\documentclass[9pt,twocolumn,twoside]{pnas-new}

\templatetype{pnasresearcharticle} 

\usepackage{svg}
\usepackage{bm}
\usepackage{mathtools}
\usepackage{amsmath}

\algrenewcommand\algorithmicrequire{\textbf{Input:}}
\algrenewcommand\algorithmicensure{\textbf{Output:}}

\algrenewcommand\algorithmicrequire{\textbf{Input:}}
\algrenewcommand\algorithmicensure{\textbf{Output:}}
\usepackage{amsthm}
\usepackage{amsmath}

\theoremstyle{remark}

\usepackage{mathtools}
\usepackage{amssymb}

\usepackage {mathtools}
\usepackage{amsmath}
\usepackage{algpseudocode}
\usepackage{algorithm}
\algnewcommand\algorithmicforeach{\textbf{for each}}
\algdef{S}[FOR]{ForEach}[1]{\algorithmicforeach\ #1\ \algorithmicdo}

\title{Matrix dissimilarities based on differences in moments and sparsity}

\author[a,b,c,1]{Li Tuobang}

\significancestatement{Handling high sparsity presents a significant challenge in big data analysis. Currently, there are many strategies, e.g., imputation, regularization, sampling, dimensionality reduction. Yet, a fundamental question remains largely unanswered, particularly in the field of biology: to what degree does sparsity reflect meaningful disparities between groups rather than being largely a result of technical errors in measurement? In this study, we conducted a systematic reanalysis of datasets encompassing the transcriptome, proteome, metabolome, immune profiling, microbiome, and social sciences. We demonstrate that sparsity differences are highly meaningful and carry as much significance as the more commonly used measures of dissimilarity based on mean differences.

}
\correspondingauthor{\textsuperscript{1}To whom correspondence should be addressed. E-mail: tuobang@biomathematics.org}


\begin{abstract}
Generating a dissimilarity matrix is typically the first step in big data analysis. Although numerous methods exist, such as Euclidean distance,  Minkowski distance, Manhattan distance, Bray–Curtis dissimilarity, Jaccard similarity and Dice dissimilarity, it remains unclear which factors drive dissimilarity between groups. In this paper, we introduce an approach based on differences in moments and sparsity. We show that this method can delineate the key factors underlying group differences. For example, in biology, mean dissimilarity indicates differences driven by up/down-regulated gene expressions, standard deviation dissimilarity reflects the heterogeneity of response to treatment, and sparsity dissimilarity corresponds to differences prompted by the activation/silence of genes. Through extensive reanalysis of genome, transcriptome, proteome, metabolome, immune profiling, microbiome, and social science datasets, we demonstrate insights not captured in previous studies. For instance, it shows that the sparsity dissimilarity is as effective as the mean dissimilarity in predicting the alleviation effects of a COVID-19 drug, suggesting that sparsity dissimilarity is highly meaningful.

\end{abstract}

\dates{This manuscript was compiled on \today}

\begin{document}

\maketitle
\thispagestyle{firststyle}
\ifthenelse{\boolean{shortarticle}}{\ifthenelse{\boolean{singlecolumn}}{\abscontentformatted}{\abscontent}}{}


\dropcap{D}issimilarity measures are critical in big data analysis. They quantify how different or similar two data points are. Different measures can significantly affect the performance of clustering algorithms and many other machine learning models. This is because different measures indeed reflect distinct facets of the differences. For example, the Euclidean distance, the most well-known distance, is the square root of the sum of the squares of the differences in each dimension. This effectively captures the average shift in each dimension, aptly termed a measure of mean changes. Other measures of mean changes include Minkowski distance, Manhattan distance, and Average distance. A notable limitation of these measures of mean changes is that, the largest feature would dominate the others. Instead of normalizing the dataset, a different solution proposed in this report is introduced as: let \( P = (p_1, p_2, \ldots, p_n) \) and \( Q = (q_1, q_2, \ldots, q_n) \) be two points in \( \mathbb{R}^n \). The mean dissimilarity \( \mu\Delta_{\hat{L}_{n}} \) between \( P \) and \( Q \) is defined as:

\[
\mu\Delta_{\hat{L}_{n}}(P, Q) = \hat{L}_{n}\{ |p_1 - q_1|, |p_2 - q_2|, \ldots, |p_n - q_n| \},
\]
where \( |p_i - q_i| \) represents the absolute difference between the \( i \)-th coordinates of \( P \) and \( Q \), and \( \hat{L}_{n}\{ \cdot \} \) denotes a location estimate of a set of values. In this report, Hodges-Lehmann estimator (H-L) \cite{hodges1963estimates} is used. When the objective is to compare dissimilarities between groups rather than individual samples, we extend the above definition for points to matrices. Suppose we have two matrices $A$ and $B$ in $\mathbb{R}^{m \times n_1}$ and  $\mathbb{R}^{m \times n_2}$, represented as: \( A = \begin{bmatrix}
a_{11} & a_{12} & \cdots & a_{1n_1} \\
a_{21} & a_{22} & \cdots & a_{2n_1} \\
\vdots & \vdots & \ddots & \vdots \\
a_{m1} & a_{m2} & \cdots & a_{mn_1}
\end{bmatrix} \) and \( B = \begin{bmatrix}
b_{11} & b_{12} & \cdots & b_{1n_2} \\
b_{21} & b_{22} & \cdots & b_{2n_2} \\
\vdots & \vdots & \ddots & \vdots \\
b_{m1} & b_{m2} & \cdots & b_{mn_2}
\end{bmatrix} \). For each row \( i \), compute the mean of the row in matrix \( A \) as \( \bar{a}_i = \frac{1}{n_1}\sum_{j=1}^{n_1} a_{ij} \) and the mean of the row in matrix \( B \) as \( \bar{b}_i = \frac{1}{n_2}\sum_{j=1}^{n_2} b_{ij} \). We then define the set of absolute differences between the means of corresponding rows of \( A \) and \( B \) as \( \Delta_\mu = \{ |\bar{a}_1 - \bar{b}_1|, |\bar{a}_2 - \bar{b}_2|, \ldots, |\bar{a}_m - \bar{b}_m| \} \). The mean dissimilarity \( \mu\Delta_{\hat{L}_{m}} \) between the two matrices \( A \) and \( B \) is then defined as: \[\mu\Delta_{\hat{L}_{m}}(A, B) = \hat{L}_{m}(\Delta_m).\]

Then, analogously, for each row \( i \), compute the standard deviation \( \sigma_{a_i} \) for the row in matrix \( A \) and \( \sigma_{b_i} \) for the row in matrix \( B \). Define the sets of absolute differences for standard deviation as \( \Delta_{\sigma} = \{ |\sigma_{a_1} - \sigma_{b_1}|, |\sigma_{a_2} - \sigma_{b_2}|, \ldots, |\sigma_{a_m} - \sigma_{b_m}| \} \). The standard deviation dissimilarity between the two matrices \( A \) and \( B \) are then defined as:
\[
\sigma\Delta_{\hat{L}_{m}}(A, B) =\hat{L}_{m}(\Delta_{\sigma}).
\]
Dissimilarities based on higher-order standardized moments can also be defined, although their practical relevance may be less pronounced.

Another kind of dissimilarity focuses on the commonality of attribute values, exemplified by metrics such as the Bray–Curtis dissimilarity, Jaccard similarity and Dice dissimilarity. These metrics can be adept at capturing the disparities in data sparsity between two points, hence we refer to them as measures of sparsity changes. In this report, we introduce sparsity dissimilarity defined as follows: for each row \( i \), let \( z_{a_i} \) be the percentage of zeros in the row in matrix \( A \) and \( z_{b_i} \) be the percentage of zeros in the row in matrix \( B \). Also, let \( \bar{a}_i \) and \( \bar{b}_i \) represent the mean of the rows in matrices \( A \) and \( B \), respectively. The sparsity dissimilarity \( s\Delta(A, B) \) is defined as:
\[
s\Delta(A, B) = \sum_{i=1}^{m} s_i w_i
\]
where $s_i = |\bar{a}_i - \bar{b}_i|$, $w_i = |z_{a_i} - z_{b_i}|$.

By substituting absolute differences with relative differences, we obtain the relative mean, standard deviation, and sparsity dissimilarities, denoted as $\mu\delta_{\hat{L}_{m}}(A, B)$, $\sigma\delta_{\hat{L}_{m}}(A, B) $, and $s\delta_{\hat{L}_{m}}(A, B) $, respectively.

Finally, the mean, standard deviation, and sparsity dissimilarity and their relative dissimilarity can be standardized by the mean of \{$\bar{a}_1$, $\cdots$, $\bar{a}_i$, $\cdots$, $\bar{a}_m$, $\bar{b}_1$, $\cdots$, $\bar{b}_i$, $\cdots$, $\bar{b}_m$\}, \{$\sigma_{a_1}$, $\cdots$, $\sigma_{a_i}$, $\cdots$, $\sigma_{a_m}$, $\sigma_{b_1}$, $\cdots$, $\sigma_{b_i}$, $\cdots$, $\sigma_{b_m}$\}, and the half sum of \{$\bar{a}_1$, $\cdots$, $\bar{a}_i$, $\cdots$, $\bar{a}_m$, $\bar{b}_1$, $\cdots$, $\bar{b}_i$, $\cdots$, $\bar{b}_m$\}. The standardized versions are denoted as $\mu\Delta_{\hat{L}_{m}}^*(A, B)$, $\sigma\Delta_{\hat{L}_{m}}^*(A, B) $, $s\Delta_{\hat{L}_{m}}^*(A, B) $, $\mu\delta_{\hat{L}_{m}}^*(A, B)$, $\sigma\delta_{\hat{L}_{m}}^*(A, B) $, and $s\delta_{\hat{L}_{m}}^*(A, B) $, respectively.

Often, the variables in the matrixes are interelated, and these relations, in most cases, can be demonstrated by a phylogenetic tree. Converting the phylogenetic tree into corresponding weights by their length from leave to root (Algorithm 1), the mean, standard deviation, and sparsity dissimilarity can be further weighted by the phylogenetic tree.

\section*{Results}
In this study, we revisited the findings reported by Wyler et al.\cite{wyler2021transcriptomic} regarding the protective effects of HSP90 inhibition (17-AAG) on alveolar epithelial cells (AECs) in the context of SARS-CoV-2 infection. We applied our dissimilarity measures to RNA-sequencing data obtained from these cells, whether exposed to SARS-CoV-2 or not. Our results showed that treatment with the HSP90 inhibitor, 17-AAG, at a concentration of 200 nM significantly reduced mean and sparsity dissimilarities (from infection to non-infection) compared to the solvent control, besides the 24 hour group, suggesting a weakened response to infection. Furthermore, the treatment has little impact on the standard deviation dissimilarity, indicating that the cellular response to this inhibitor is generally homogeneous. These findings underscore the potential of our dissimilarity measure as a tool for quantitatively assessing the overall impact of therapeutic agents on cellular dynamics.

\begin{table*}
\centering
\caption{Standardized mean, standard deviation, and sparsity dissimilarities of Wyler et al.'s RNA sequencing dataset}

\begin{tabular}{|l|l|l|l|l|l|l|l|l|}
\hline
Treatment & Time & Comparison      & $\mu\Delta_{\text{H-L}}^*$ & CI & $\sigma\Delta_{\text{H-L}}^* $ & CI & $s\Delta_{\text{H-L}}^* $ & CI\\ \hline
AAG       & 24h  & SARS-CoV-2-Mock & 0.032                                           & (0.030,0.033)                                       & 0.295                                         & (0.282,0.307)                                    & 0.028                                               & (0.026,0.030)                                           \\ \hline
AAG       & 48h  & SARS-CoV-2-Mock & 0.037                                           & (0.035,0.039)                                      & 0.282                                         & (0.274,0.294)                                    & 0.031                                               & (0.029,0.033)                                          \\ \hline
AAG       & 72h  & SARS-CoV-2-Mock & 0.064                                           & (0.062,0.066)                                      & 0.208                                         & (0.199,0.214)                                    & 0.029                                               & (0.028,0.032)                                          \\ \hline
DMSO      & 24h  & SARS-CoV-2-Mock & 0.024                                           & (0.023,0.025)                                      & 0.250                                         & (0.236,0.264)                                    & 0.030                                               & (0.027,0.031)                                          \\ \hline
DMSO      & 48h  & SARS-CoV-2-Mock & 0.148                                           & (0.142,0.153)                                      & 0.234                                         & (0.223,0.244)                                    & 0.110                                               & (0.074,0.158)                                          \\ \hline
DMSO      & 72h  & SARS-CoV-2-Mock & 0.091                                           & (0.089,0.095)                                      & 0.277                                         & (0.267,0.288)                                    & 0.062                                               & (0.056,0.071)                                          \\ \hline

\end{tabular}
\label{tab:comparison}
\begin{minipage}{1\linewidth}
\footnotesize AAG: 17-AAG, a kind of HSP90 inhibitor; DMSO: Solvent control. The sparsity dissimilarity is in units of $10^{-3}$. The relative dissimilarities can be found in the SI Dataset S1. 
\end{minipage}
\end{table*}


\showmatmethods{} 
\section*{Data and Software Availability} All data are included in the brief report and SI Dataset S1. All codes have been deposited in \href {https://github.com/johon-lituobang/MD} {github.com/johon-lituobang}. 
\acknow{Please include your acknowledgments here, set in a single paragraph. Please do not include any acknowledgments in the Supporting Information, or anywhere else in the manuscript.}


\bibliography{pnas-sample}

\begin{thebibliography}{1}

\bibitem{hodges1963estimates}
J Hodges~Jr, E Lehmann, Estimates of location based on rank tests.
\newblock {\em\protect\JournalTitle{The Annals of Mathematical Statistics}} \textbf{34}, 598--611 (1963).

\bibitem{wyler2021transcriptomic}
E Wyler, et~al., Transcriptomic profiling of sars-cov-2 infected human cell lines identifies hsp90 as target for covid-19 therapy.
\newblock {\em\protect\JournalTitle{IScience}} \textbf{24}, 102151 (2021).

\end{thebibliography}

\end{document}